# Radar Cross Section Reduction of Microstrip Patch Antenna using Metamaterial Techniques

Syamly S. B, Job Chunkath

APJ Abdul Kalam Technological University, Thiruvananthapuram
Department of Electronics and Communication Engineering,
Government Engineering College Thrissur, Kerala, India
E-Mail: jobchunkath@gmail.com

*Abstract* – *Radar cross section (RCS) reduction has become one of the critical research areas in recent years. The RCS of the target should be small to avoid detection. Different methods are used to reduce RCS, but the major challenge with many RCS minimization methodologies is that, it may deteriorate some antenna parameters. When antenna mode RCS is considered; structural mode RCS, and antenna parameters are critical, as the structure should be an antenna and a RCS reducing structure simultaneously. The techniques like applying Radar Absorption Material (RAM) entirely over the target, deployment of Energy Band Gap (EBG) structures, the use of passive, active cancellation, and polarization conversion are prevalent methods to reduce RCS. The manifestation of metamaterial property in an antenna results in the antenna's electromagnetic characteristics becoming negative for a particular bandwidth. Thus the RCS of the antenna can be reduced to a minimum range by loading the metamaterial structures. This paper discusses the application of polarization conversion method (PCM), L- structured and Square - structured fractal metamaterial antenna for RCS reduction. This paper reports the simulation, fabrication, and testing of the above antennas with their performance comparison. The antennas are designed for 4.3GHz frequency with a total dimension of 80mm×80mm×1.6mm. Antenna parameters like return loss, gain, radiation pattern, and bandwidth are analyzed along with the RCS. The L- structured metamaterial antenna implemented has a 29.37% larger bandwidth than the reference patch antenna with a gain of 2.94dB with a return loss of -28.28dB.*

## I. INTRODUCTION

The investigations on Radar Cross Section (RCS) reduction method is rapidly establishing into a well-known research area. In the military application, RCS reduction has received significant attention. Radar cross section (RCS) reduction of an antenna has gained more attention in recent years due to the increased reliance on communication for airborne systems. This is due to the increased use of antennas which also causes scattering of incident radiation. Thus almost all systems, whether it is civilian or military, require antennas with low RCS to avoid detection.

There are different methods conventionally used for RCS reduction, which is based on stealth shaping, use of low-observable Radar Absorbing Material (RAM), bionics principle [13], biased ferrite substrate [1], Energy Band Gap (EBG) structure, and Frequency Selective Structure (FSS). FSS and EBG structures are commonly used nowadays to reduce RCS. The antenna polarization can also be utilized for reducing RCS. All methods result in better RCS reduction but lead to performance degradation of some antenna parameters.

At present, the focus of the research is on microstrip patch antenna, because of its low profile characteristics. The latest research in metamaterials has resulted in a more innovative antenna design. The metamaterial structure used in antenna reduces the antenna RCS because it reduces antenna size, increases its efficiency, reduces surface waves and mutual coupling between antenna elements. It also eliminates grating lobes in antenna arrays and leads to improve the overall antenna performance.

The RCS reduction of the antenna has gained extensive research interest due to the development of electronic countermeasures technology for the military. Over the past decades, many solutions have been presented to reduce RCS. All of these methods help to decrease RCS, but often the antenna performance gets degraded while minimizing RCS. Ideally, the antenna gain, efficiency, directivity, and reflection coefficient should be maintained along with low RCS. The RCS is expressed in decibel square meter (dBsm).

In this paper, new structures for metamaterial patch antennas are proposed. These have a similar planar size as that of the conventional reference patch antenna but have different metamaterial structures. The analyses of these antennas have demonstrated an improvement in the antenna parameters and RCS reduction. The gain and bandwidth of square-structured fractal antenna increased to 3.2dB and 201.9MHz, respectively. A return loss of -28.28dB and RCS reduction of -36.71dBsm, was achieved by the proposed L-structured metamaterial antenna.

The paper describes the antenna design using metamaterial structures to obtain low RCS and return loss for microstrip patch antennas. This paper is organized as follows; Section II describes different Radar Cross Section Reduction Methods. In Section III, the Proposed Antenna Design is discussed. Section IV details the Fabrication & Analysis of these antennas. A discussion on Results and Conclusion are included in Section V and Section VI, respectively.

## II. RADAR CROSS SECTION REDUCTION METHODS

The major factors that affect the RCS reduction are, the material used in target fabrication, Radar transmitter frequency, the direction of illuminating Radar, incident angle, reflected angle, polarization, and airframe physical geometry. RCS reduction is mainly investigated in military applications. The targets can be military aircraft, missiles, and ships. These targets are manufactured using different kinds of materials.

The Radar Absorption Materials (RAM) is mainly used in stealth technology to avoid vehicles or structure detection by Radar. The RAM is designed and shaped in such a way as to absorb the incident Radio Frequency (RF) energy [2].These materials are designed to resist the reflection or transmission of electromagnetic radiations. There are resonant and broadband absorbers. The resonant absorbers utilize the resonant property of the material to function and its effectiveness depends on the frequency of the incident radiation.

The broadband absorber soaks up the incident energy that comes from the Radar [3]. The incident RF energy is converted into heat. Thus it decreases the energy scattered or reflected towards the Radar. The RAM is designed and shaped in such a way as to absorb the incident Radio Frequency (RF) energy and reduce RCS. The absorption level depends upon the frequency in which the Radar is operated. A significant factor deciding the magnitude of the radiation absorbed is the composition of the material used, no composition available can absorb the complete range of Radar frequencies. Thus RAM does not provide invisibility but lowers the Radar cross-section for a specific range of frequencies.

The military vehicles, such as ships or aircraft, size cannot be changed beyond their operational capabilities. Hence, the cross-section of the system cannot be decreased beyond a limit. The only way is to reduce other parameters which cause the RCS of the system. The shape of the target can be changed so that the scattered energy can be reflected in a direction away from the Radar. For example, a replacement of a flat surface with a curved surface can minimize narrow and intense specular lobes.

The shaping method is usually challenging to exploit or expensive to implement for vehicles or objects in service [3]. Also, if the objects are not electrically large, then shaping is not very useful.

$$\text{Overall RCS}, \sigma = \left|\sqrt{\sigma_s} + \sqrt{\sigma_a}e^{j\emptyset}\right|^2 \quad (1)$$

Where,
$\emptyset$ = phase difference between two modes.
$\sigma_s$ = structural mode RCS.
$\sigma_a$ = antenna mode RCS.

When electromagnetic waves are incident on the antenna surface, some energy is scattered back to space, this is called structural mode scattering. Due to the antenna effect remaining part of the energy is absorbed, and some part of the absorbed energy is again scattered into space due to impedance mismatches. The scattering mentioned above is called antenna mode scattering. The RCS due to these scattering is, respectively, known as structural mode RCS and antenna mode RCS. The expression for overall RCS is given by Eqn. (1).

A method of using Energy Band Gap (EBG) structures to minimize antenna RCS is described in [4]. The EBG structures act as an artificial magnetic conductor (AMC) in a frequency band. The AMC has a reflection coefficient of +1 and -1 for the perfect electric conductor (PEC). It is observed that it has low in-band RCS. When these two conductors are combined, the reflected waves get canceled, and thus the backward scattering is decreased. In the X-band of frequencies, the EBG structure reduces the RCS while the antenna array operates in the S-band of frequencies.

In [4], it is seen that RCS is diminished using EBG structures shaped like a mushroom. This is realized by an array of metallic patches with periodic structures connected with the ground plane. The microstrip patch antennas make use of this kind of EBG structure for making low-profile antennas with improved performance. These are also used in-phase reflection bandgap and surface-wave rejection bandgap. The method of reducing RCS of the patch array antenna at in-band and out-of-band frequencies using mushroom-like EBG structures are discussed in [5].

A performance improvement compared to all the above methods is achieved, with the advent of the polarization conversion technique. In papers [6], [7], and [8] discuss the effect of the polarization method. It is observed using the polarization effect, the RCS minimization and antenna performance can be attained as per the requirement. In [7], the energy bandgap (EBG) structure along with the polarization method is used. The polarization rotation mechanism along with passive EBG cancellation is utilized in [8]. In [9], the method describes a flexible cylindrically curved ground plane on which an AMC structure is created to lower the RCS. The magnetic absorption material proposed in [10] decreases RCS but has a narrow bandwidth.

A metamaterial structure has multiple similar units, which can alter the antenna parameters. Hence, the free space impedance matching with impedance due to the electric and magnetic component of antenna can be achieved, thus resulting in minimum scattering of the electromagnetic waves at the interface.

The metamaterials are implemented from ordinary materials by designing different shapes, so they function as an artificial material with parameters that can be used in improving antenna performance. The magnitudes, phase, and polarization of electromagnetic waves are controlled in the metamaterial [11]. Different methods of metamaterial loading techniques like superstrate loading [16], coplanar loading, and cavity ground loading are explained in [11]. This paper discusses enhancing the RCSR (Radar Cross Section Reduction) and gains through a Chessboard-Like Metamaterial Surface (CLMS). It is observed that CLMS cancels the back-scattering effect caused by the AMC phase difference [14]. The RCS is reduced by metamaterial structure introduced in the antenna, instead of a single patch and thus improves the antenna performance.

### III. PROPOSED ANTENNA DESIGN

The most commonly used model for RCS reduction is the Polarization Conversion Metamaterial (PCM) structure, this is a chessboard-like structure having four sections that are separated by a distance. In this method polarization conversion metamaterial develops a polarization rotation mechanism, which results in RCSR. The RCSR is achieved when the polarization of reflected waves gets controlled [6].

The antenna structure has a dimension of 80mm× 80mm using Flame Retardant (FR4) material as the substrate. The FR4 material has a permittivity of 4.4 ($\varepsilon_r$ = 4.4), with a loss tangent of 0.025 and 2mm thickness. A coaxial feeding is used for signal coupling to the antenna. Coaxial feeding is used to decrease the chance of spurious radiation, and also it is easy to fabricate and match with the antenna, but it is difficult in the case of thick substrates. The structural dimensions of designed antennas are mentioned in Table 1.

Table 1: The dimension of the antennas implemented

| Parameter | Antenna Structure Dimensions (mm) | | |
|---|---|---|---|
| | Antenna A | Antenna B | Antenna C |
| Center Patch Width | 16 | 16 | 16 |
| Center Patch Length | 15.6 | 14.5 | 14.8 |
| Ground Width | 80 | 80 | 80 |
| Ground Length | 80 | 80 | 80 |
| Thickness | 1.6 | 1.6 | 1.6 |

A: PCM structured antenna, B: L - structured metamaterial antenna, C: Square - structured fractal metamaterial antenna.

The antenna is designed to operate with a resonant frequency of 4.3GHz. As per the simulation, the antenna operates at 4.3GHz without any shift. The antenna is loaded with a metamaterial structure. This has properties like negative relative permittivity ($\varepsilon_r$), relative permeability ($\mu_r$), and negative refractive index. The properties are analyzed using Ansys® High-Frequency Structure Simulator (HFSS) software and MATLAB®.

The PCM unit cells are arranged around the radiating patch. Coaxial feeding is given to this radiating patch. The four sections in the antenna contribute reflections, and adjacent sections have reflections with a 180° phase shift, which cancels each other. This cancellation results in RCS reduction while enhancing other antenna parameters.

*A. Microstrip Patch Antenna Design*

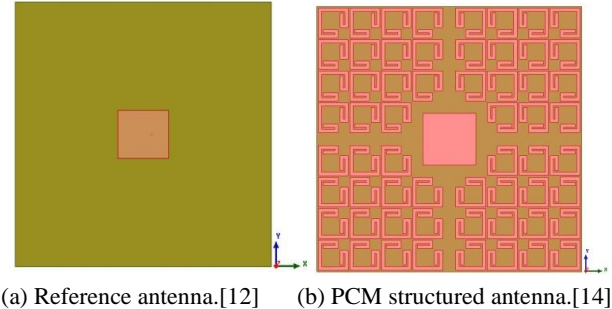

(a) Reference antenna.[12]     (b) PCM structured antenna.[14]
Fig. 1: Antenna Structures

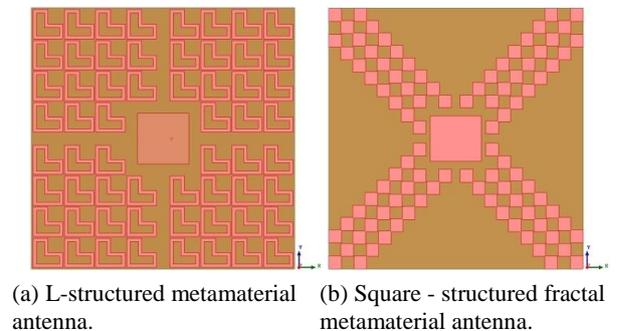

(a) L-structured metamaterial antenna.     (b) Square - structured fractal metamaterial antenna.

Fig. 2: Proposed Antenna Structures [11]

The primary stage of designing an antenna is based on the equations from [12], and then dimension analysis is done. A reference antenna, as shown in Fig. 1(a), consists of a center patch. The center patch dimension is designed first, and all the parameters were analyzed. The antenna is designed with a resonant frequency of 4.3GHz on the FR4 substrate with relative permittivity of 4.4 with a thickness of 1.6mm. The first step is to design a single element and then do the analysis. The radiation pattern, RCS, return loss, and gain of the reference antenna are obtained from Ansys® HFSS and shown in Table 2.

The proposed antenna structures are shown in Fig. 1(b), Fig.2 (a), and Fig.2 (b). Each antenna consists of a center patch, ground, substrate, metamaterial structure, and feed port. The proposed antennas were designed with similar dimensions and the parameters attained are investigated. All these antennas are fed using the coaxial feeding technique which produces low spurious radiation [15]. Coaxial feeding is done by a matching 50Ω impedance adapter.

In an attempt to improve antenna performance, the metamaterial is loaded in all three structures. It is observed that the metamaterial structure reduced the RCS value to a great extent in all the cases by canceling the scattering due to the incident wave.

## IV. FABRICATION & ANALYSIS

The antennas investigated are fabricated and fed by a coaxial feeding technique. A microstrip patch antenna consists of ground, radiating patch, and a substrate. The PCM structured antenna is shown in Fig. 3. Fig. 3(a) shows the patch and Fig. 3(b) is the ground structure with adaptor for coaxial feed.

The L - structured metamaterial antenna's patch is shown in Fig. 4(a) and the ground structure with feed is depicted in Fig. 4(b). In Fig. 5 (a) and Fig. 5(b) the patch and ground structure with the feed of Square structured fractal metamaterial antenna is displayed.

The antenna parameters such as gain, bandwidth, and radiation pattern are measured. The return loss of the antennas is measured using a Vector Network Analyzer (VNA) and the radiation measurements were carried out inside an anechoic chamber.

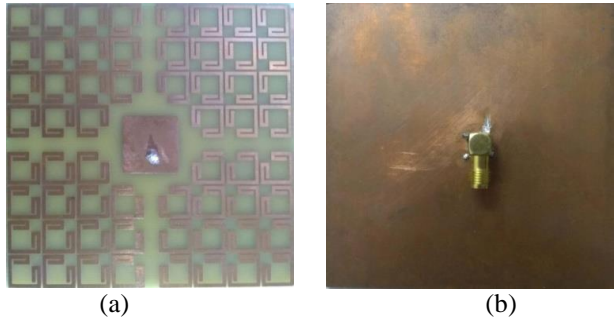

(a)                      (b)

Fig. 3: PCM structured antenna.

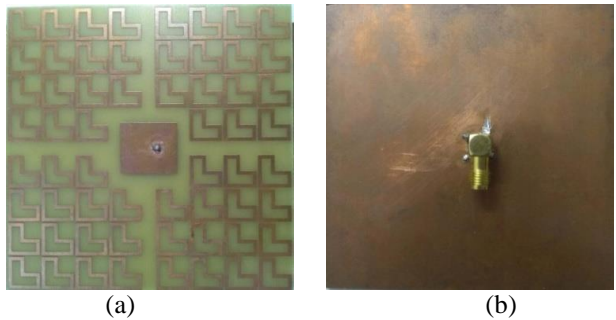

(a)                      (b)

Fig. 4: L - structured metamaterial antenna.

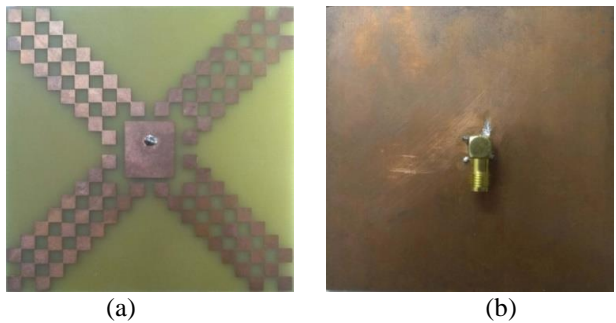

(a)                      (b)

Fig. 5: Square structured fractal metamaterial antenna.

## V. RESULTS

The Ansys® HFSS software is used to design and simulate the antenna. Among the available simulation software, Ansys® HFSS is chosen because it gives a result that is very close to the experimental results. The antenna is designed for a center frequency of 4.3GHz. The antenna is designed for reducing the Radar cross section (RCS) and also to achieve a return loss ($S_{11}$ parameter) of less than -10dB at the center frequency.

In HFSS, all the antenna parameters are obtained and validated. An analysis is done based on comparing three antennas. The main feature of these antennas is that they are loaded with a metamaterial structure. The overall structure and dimensions of these antennas are similar. The metamaterial properties of these antennas are analyzed.

All three antennas have negative electromagnetic characteristics. This means that they have negative relative permittivity, permeability, and refractive index. These characteristics obtained through simulation are verified with the help of VNA and an anechoic chamber.

### A. Reference Patch Antenna Characteristics

A simulation of the reference patch antenna design yields a bandwidth of 143MHz with a gain of 2.8dB. The antenna has an RCS value of -27.98dB and a Return loss ($S_{11}$) of -16.88dB. These values are shown in the first row of Table 2. From these values, it is clear that the antenna with a simple patch structure cannot provide adequate Radar cross section reduction, return loss, and gain. The return loss and RCS values are plotted in Fig.6 and Fig.7, respectively.

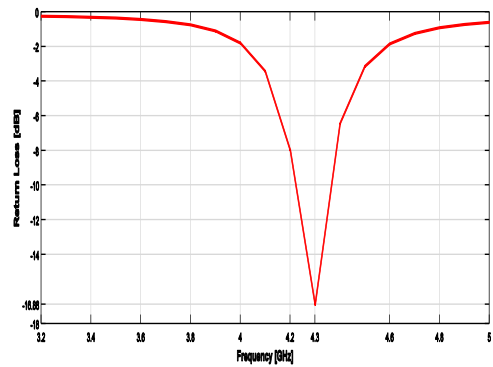

Fig. 6: Return loss of reference patch antenna.

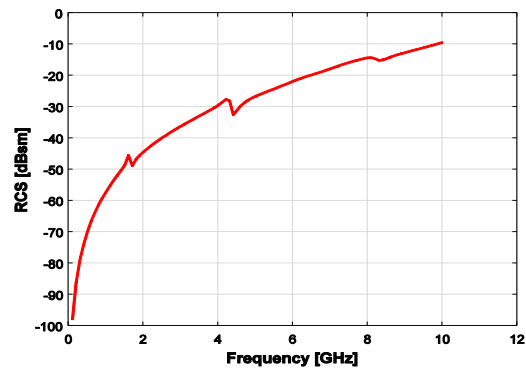

Fig. 7: Radar cross section of reference patch antenna.

### B. Characteristics of PCM Structured Antenna

The PCM structured antenna is realized on an FR4 substrate with a thickness of 1.6mm. The antenna is loaded with a metamaterial structure around the center patch. The results obtained after testing for the return loss, radiation pattern, and RCS plot is shown in Fig. 8, and Fig. 9, respectively.

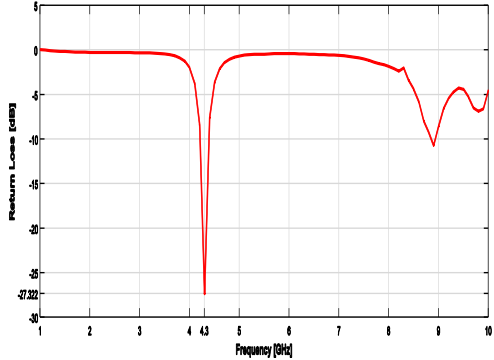

Fig. 8: Return Loss of PCM Structured antenna.

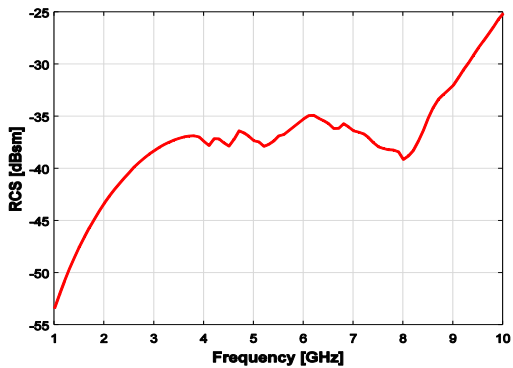

Fig. 9: Radar Cross Section of PCM Structured antenna.

### C. L-Structured Metamaterial Antenna Characteristics

The return loss, radiation pattern, and RCS plot obtained from test results of L-structured metamaterial antenna are given in Fig. 10 and Fig. 11, respectively.

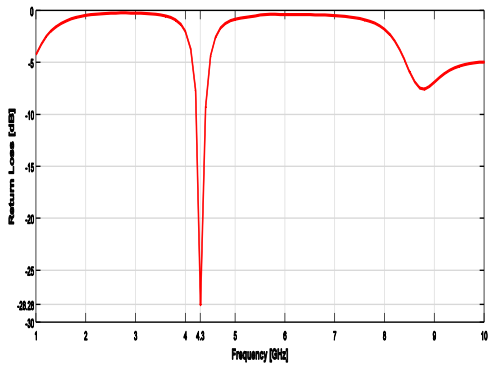

Fig.10: Return Loss of L-structured metamaterial antenna.

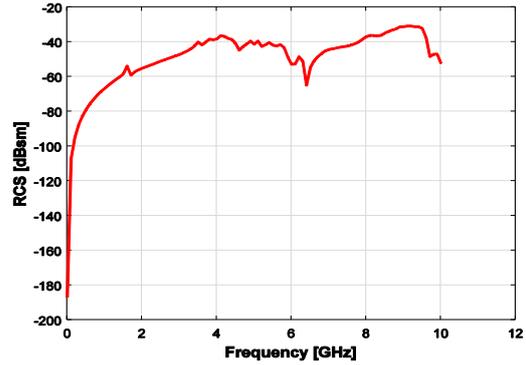

Fig. 11: Radar Cross Section of L-structured metamaterial antenna.

### D. Square - Structured Fractal Metamaterial Antenna. Characteristics

The test results of Square - structured fractal metamaterial antenna are plotted in Fig. 12 and Fig. 13, respectively.

The test results of the three antenna structures fabricated are plotted. The comparison of values with the reference patch antenna shows that the return loss has a significant improvement while reducing the RCS. The bandwidth of the antennas has improved, while the gain has only increased marginally. The detailed results are given in Table 2.

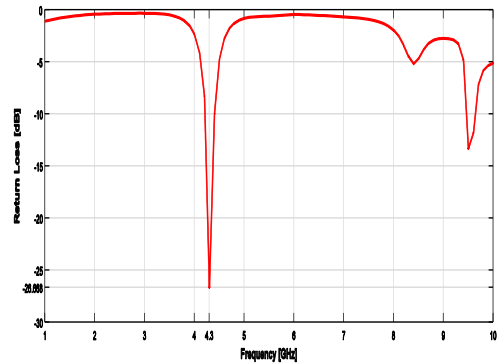

Fig. 12: Return Loss of square-structured fractal metamaterial antenna.

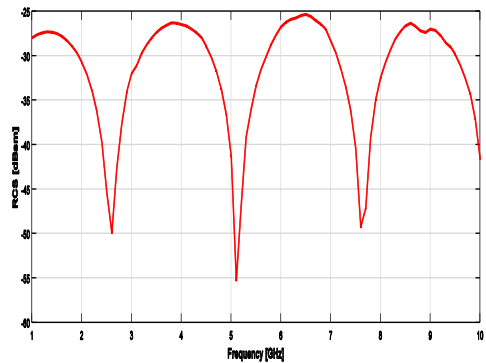

Fig. 13: Radar Cross Section of square-structured fractal metamaterial antenna.

JEEE, Volume 14, Number 2, October 2021
Table 2: Performance comparison of different antennas

| Sl. No. | Antenna | Antenna Bandwidth (MHz) | Gain (dB) | RCS Value (dBsm) | Return Loss (dB) $S_{11}$ |
|---|---|---|---|---|---|
| 1. | A | 143MHz | 2.8 | -27.98 | -16.88 |
| 2. | B | 179MHz | 2.9 | -37.14 | -27.322 |
| 3. | C | 185MHz | 2.94 | -36.71 | -28.28 |
| 4. | D | 201.9MHz | 3.2 | -27.36 | -26.667 |

A: Reference antenna, B: PCM structured antenna, C: L - structured metamaterial antenna, D: Square - structured fractal metamaterial antenna.

## VI. CONCLUSION

An effective method for Radar cross section reduction is vital in the communication system and defense applications. The suitability of three different techniques for RCS reduction was investigated in this paper. It is observed that each method has its advantages as well as disadvantages. A single method capable of achieving objectives, like low return loss, high gain, and high bandwidth along with good RCS minimization was difficult to realize.

The performance evaluation of three antennas with RCS reduction methods compared in this paper has the following significant results,

- The metamaterial loading has resulted in the decrease of Radar cross-section in all three fabricated antennas.
- A sharp decrease in return loss was achieved by all the antenna designs in comparison with the reference patch antenna.
- A significant increase in bandwidth was reported for all the antennas with a marginal increase in gain.
- The PCM structured antenna achieved the best RCS value of -37.14dBsm.
- The lowest return loss of -28.28dB was achieved by an L-structured metamaterial antenna.
- The highest bandwidth of 201.9MHz was attained with a Square-structured fractal metamaterial antenna.

Thus it can be inferred that the metamaterial design imparts an overall improvement in antenna parameters while achieving RCS reduction.

## ACKNOWLEDGMENT

The authors would like to thank the Head, faculty, and staff associated with the Department of Electronics at Cochin University of Science and Technology (CUSAT) for providing the laboratory facilities for the antenna characterization.
## REFERENCES

[1] D. M. Pozar, "RCS Reduction for a Microstrip Antenna Using a Normally Biased Ferrite Substrate," IEEE Microwave and Guided Wave Letters, vol. 2, no. 5. pp. 196–198, 1992.
[2] A. Shater and D. Zarifi, "Radar Cross Section Reduction of Microstrip Antenna Using Dual-Band Metamaterial Absorber," vol. 32, no. 2, pp. 135–140, Feb. 2017.
[3] M. I. Skolnik, Radar Handbook, 3rd edition. McGraw-Hill, 2008.
[4] J. Zhang, J. Wang, M. Chen, and Z. Zhang, "RCS reduction of patch array antenna by electromagnetic band-gap structure," IEEE Antennas Wirel. Propag. Lett., vol. 11, pp. 1048–1051, 2012.
[5] S. G. Mao Long, Wen Jiang, "Double-layer miniaturised-element metasurface for RCS reduction," IET Microwaves, Antennas Propag., vol. 11, no. 5, pp. 705–710, 2017.
[6] Y. Liu, Y. Hao, K. Li, and S. Gong, "Radar cross section reduction of a microstrip antenna based on polarization conversion metamaterial," IEEE Antennas Wirel. Propag. Lett., vol. 15, pp. 80–83, 2016.
[7] K. Li, Y. Liu, Y. Jia, and Y. J. Guo, "A circularly polarized high-gain antenna with low RCS over a wideband using chessboard polarization conversion metasurfaces," IEEE Trans. Antennas Propag., vol. 65, no. 8, pp. 4288–4292, 2017.
[8] Y. Jia, Y. Liu, Y. J. Guo, K. Li, and S. Gong, "A dual-patch polarization rotation reflective surface and its application to ultra-wideband RCS reduction," IEEE Trans. Antennas Propag., vol. 65, no. 6, pp. 3291–3295, Jun. 2017.
[9] C. A. Balanis, W. Chen, C. R. Birtcher, and A. Y. Modi, "Cylindrically curved checkerboard surfaces for Radar cross section reduction," IEEE Antennas Wirel. Propag. Lett., vol. 17, no. 2, pp. 343–346, Feb. 2018.
[10] H. B. Baskey, E. Johari, and M. J. Akhtar, "Metamaterial structure integrated with a dielectric absorber for wideband reduction of antennas Radar cross section," IEEE Trans. on Electromagn. Compat., vol. 59, no. 4, pp. 1060–1069, 2017.
[11] Y.-J. Zheng et al., "Metamaterial-based patch antenna with wideband RCS reduction and gain enhancement using improved loading method," IET Microwaves, Antennas Propag., vol. 11, no. 9, pp. 1183–1189, 2017.
[12] Constantine A. Balanis, Antenna theory: analysis and design, 3rd edition. John Wiley & Sons, Inc., 2005.
[13] W. Jiang, Y. Liu, S. Gong, and T. Hong, "Application of bionics in antenna Radar cross section reduction," IEEE Antennas Wirel. Propag. Lett., vol. 8, pp. 1275–1278, 2009.
[14] H. Jiang, Z. Xue, Q. Zeng, W. Li, and W. Ren, "High-gain low-RCS slot antenna array based on checkerboard surface," IET Microwaves, Antennas Propag., vol. 12, no. 2, pp. 237–240, 2018.
[15] C. Wu, K. -L. Wu, Z. Bi and J. Litva, "Modelling of coaxial-fed microstrip patch antenna by finite difference time domain method," in Electronics Letters, vol. 27, no. 19, pp. 1691-1692, 12 Sept. 1991.
[16] Y. Zheng et al., "Wideband gain enhancement and RCS reduction of Fabry–Perot resonator antenna with chessboard arranged metamaterial superstrate," IEEE Trans. Antennas Propag., vol. 66, no. 2, pp. 590–599, 2018.
64